# Finite-time Thin Film Rupture Driven by Generalized Evaporative Loss

Hangjie Ji, Thomas P. Witelski

*Department of Mathematics, Duke University*

December 14, 2015

**Abstract**

Rupture is a nonlinear instability resulting in a finite-time singularity as a fluid layer approaches zero thickness at a point. We study the dynamics of rupture in a generalized mathematical model of thin films of viscous fluids with evaporative effects. The governing lubrication model is a fourth-order nonlinear parabolic partial differential equation with a non-conservative loss term due to evaporation. Several different types of finite-time singularities are observed due to balances between evaporation and surface tension or intermolecular forces. Non-self-similar behavior and two classes of self-similar rupture solutions are analyzed and validated against high resolution PDE simulations.

## 1. Introduction

In coating flows at small scales, used in applications like microfluidics and painting, surface tension plays a strong role in the dynamics of thin layers of viscous fluids spreading over solid substrates [39, 12]. For this class of problems, the governing equations of fluid dynamics can be reduced to a single evolution equation, called the Reynolds lubrication equation, for the evolution of the thickness (or height $h$ of the free-surface) of the fluid layer [41, 44, 16].

The behavior of the fluid will be strongly affected by wetting properties of the substrate, namely whether the solid attracts and encourages the spreading (called a hydrophilic or wetting material), or repels the fluid (called a hydrophobic or non-wetting material). One means of modeling this is through adding a contribution to the dynamic pressure representing molecular interactions between the fluid and the solid, generally called the disjoining pressure $\Pi(h)$. Consequently, the Reynolds equation for such one-dimensional models of coating flows takes the form of a fourth-order nonlinear partial differential equation for $h(x,t)$,

$$\frac{\partial h}{\partial t} = \frac{\partial}{\partial x}\left(h^3 \frac{\partial}{\partial x}\left[\Pi(h) - \frac{\partial^2 h}{\partial x^2}\right]\right), \tag{1}$$

where the influence of surface tension is manifested through the linearized curvature of the free-surface, $h_{xx}$. This equation is derived using the lubrication approximation from the classic Navier Stokes equations in the limit of low Reynolds number. The disjoining pressure characterizes key material properties in the model and qualitatively changes solution behaviors. For hydrophobic substrates, the disjoining pressure will act to oppose the diffusive spreading driven by surface tension and can generate instabilities to uniform coatings. For ideal hydrophobic materials, the simplest model for the disjoining pressure is $\Pi(h) = A/h^3$ [31], where $A$ is called a Hamaker constant. For this case, early studies demonstrated this model has instabilities in the film thickness [51]. These instabilities lead to rupture at a finite-time, $t = t_c$, with $h \to 0$ at an isolated point, $x = x_c$ and the nonlinear dynamics were shown to be given by a self-similar

*Email address:* `hangjie@math.duke.edu`   (919)-937-4606 (Hangjie Ji)

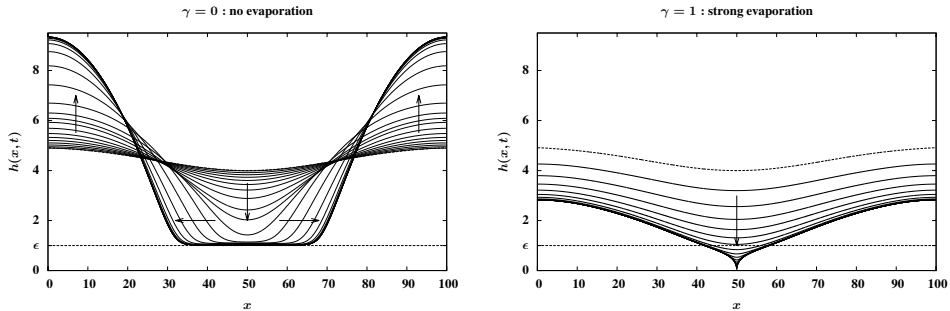

Figure 1: A comparison of numerical simulations starting from identical initial conditions, $h_0(x) = 5 - \exp(-0.001(x-50)^2)$, and system parameters, $P_0 = -1$, $K_0 = 0.1$: (left) without evaporation, for equation (1) (equivalent to model (3) with $\gamma = 0$) and (right) with strong evaporation, model (3) with $\gamma = 1$. Without evaporation, dewetting is evident with a lower bound on $h$ for all times (left), while setting $\gamma = 1$ yields finite-time rupture (right).

solution [54, 53],

$$h(x,t) \sim \tau^{1/5} H(\eta), \qquad \tau = t_c - t, \qquad \eta = \frac{x - x_c}{\tau^{2/5}}, \qquad \text{as } t \to t_c. \tag{2}$$

This model is problematic since the solution cannot be continued past the time of the first rupture, however, this can be avoided by considering more detailed models for $\Pi(h)$. In [11], conditions on the form of $\Pi(h)$ were determined so that the solutions of (1) remain positive for all times, $\Pi(h) = Ah^{-3}(1 - \epsilon/h)$ is a simple example that includes both attractive van der Waals forces and short range repulsive forces (Born repulsion). Physically, there is a film thickness $h = O(\epsilon) > 0$ set by the intermolecular forces that acts as a lower bound [48]. The solutions follow (2) until the minimum thickness approaches $h_{\min} = O(\epsilon)$, thereafter, that minimum will spread to form a growing "dry spot" while the bulk of the fluid moves to form droplets [28], see Figure 1(left). The film breakup, the development of dry spot and further morphological changes is usually called "dewetting".

The dynamics of thin films subject to fluid evaporation and vapor condensation are also interesting and challenging with applications to precorneal tear film [4], thermal management [38] and environmental control. Several models [1, 2, 13] have been constructed to characterize the evaporating/condensing liquid films. Burelbach et al. [13] first proposed a one-sided model to describe the dynamics of the liquid decoupled from the dynamics of the surrounding vapor. Based on this model, Oron and Bankoff [43, 42] studied the dynamics of a condensing thin liquid film but neglected effects like thermocapillarity and vapor thrust. A new evaporation model was derived by Ajaev [2, 1] that incorporates thermal effects, surface tension and disjoining pressure. In addition, a two-sided model was introduced by Sultan et al. [49] which considers the influence of both vapor and liquid phases. For a thorough discussion on the modeling and numerical studies of evaporating thin films, see [16]. Various forms of evaporation loss or condensation source terms have been used in the literature. For instance, the standard one-sided model [13] and two-sided model [49] only include thermal effects in the mass flux through the interface.

Rupture and long-time dynamics of dewetting in (1) have attracted extensive applied interest, and some rigorous analyses have also been developed [34, 15]. While (1) conserves fluid mass for all times and describes non-volatile liquids, what has not been studied to the same degree is whether for volatile liquids, there are mathematical forms of evaporative loss terms that can significantly change the dynamics. Although previous studies [11] have analytically verified that the disjoining pressure involved in (1) precludes the film from rupturing, the positivity of solutions cannot be guaranteed with instabilities enhanced by the evaporative effects. In particular, Fig. 1(right) shows that the evaporative flux can overcome the influence of the regularized disjoining pressure to yield finite-time rupture singularities; this is the focus of our work.

Motivated by the model of thin films with evaporation/condensation effects from the work of Ajaev [1], we explore the behavior of a lubrication model on a finite domain, $0 \leq x \leq L$,

$$\frac{\partial h}{\partial t} = \frac{\partial}{\partial x}\left(h^3 \frac{\partial p}{\partial x}\right) - J, \tag{3a}$$

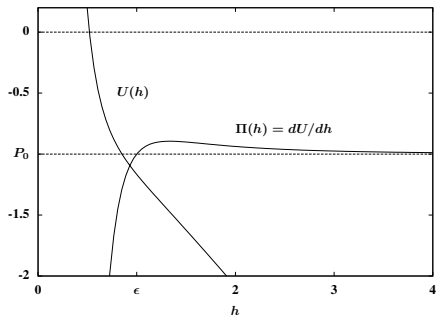

Figure 2: A sketch of the pressure $\Pi(h)$ in (3d) and the corresponding generalized potential $U(h)$ with $P_0 = -1$.

where the evaporative loss is

$$J(h) \equiv -\frac{\gamma p(h)}{h + K_0}, \tag{3b}$$

where $K_0$ is called the evaporative kinetic parameter [2], $\gamma$ is a scaling constant, and the pressure is defined as

$$p(h) \equiv \Pi(h) - h_{xx}. \tag{3c}$$

For convenience, we will generalize $\Pi(h)$ to include a constant $P_0$ corresponding to a vapor pressure [45],

$$\Pi(h) = \frac{A}{h^3}\left(1 - \frac{\epsilon}{h}\right) + P_0, \tag{3d}$$

where $\epsilon > 0$ characterizes the scale of the ultrathin film where the intermolecular forces are dominant. Our evaporative loss $J(h)$ combines the evaporative flux from Ajaev [1] and a disjoining pressure that accounts for both van der Waals interactions and Born repulsion. In this paper, we set $A = 1$ and $\epsilon = 1$, letting $h$ be normalized with respect to the van der Waals film thickness scale. Starting from a given initial condition $h_0(x)$ at $t = 0$, the dynamics will be subject to no-flux and normal-contact boundary conditions at the edges of the domain,

$$p_x(0,t) = p_x(L,t) = 0, \qquad h_x(0,t) = h_x(L,t) = 0, \tag{3e}$$

which are equivalent to specifying homogeneous Neumann conditions $h_x = h_{xxx} = 0$.

It will be useful to define the potential $U(h)$ as the integral of $\Pi(h)$,

$$U(h) = \int \Pi(h)\, dh. \tag{4}$$

For $\Pi(h)$ defined by (3d), the potential is $U(h) = -\frac{1}{2h^2} + \frac{1}{3h^3} + P_0 h$. A typical plot of this $U(h)$ and the corresponding $\Pi(h)$ with $P_0 < 0$ in Figure 2 shows the structure of the generalized potential and pressure. The disjoining pressure $\Pi(h)$ has a unique maximum at $h_{\text{peak}} > 1$,

$$\lim_{h \to 0} \Pi(h) = -\infty, \qquad \lim_{h \to \infty} \Pi(h) = P_0, \qquad \lim_{h \to 0} U(h) = \infty, \qquad \lim_{h \to \infty} U(h) = -\infty.$$

It is worth mentioning that unlike the potential used in [29], the generalized potential $U(h)$ does not have a local minimum for $P_0 < 0$.

Total mass of the film is of fundamental interest and its rate of change can be obtained from integrating (3a) and applying the boundary conditions,

$$M(t) = \int_0^L h\, dx, \qquad \frac{dM}{dt} = \gamma \int_0^L \frac{p}{h + K_0}\, dx. \tag{5}$$

Using (3c) and integration by parts, this can be expanded as

$$\frac{dM}{dt} = -\gamma\left(\int_0^L \frac{h_x^2}{(h + K_0)^2}\, dx - \int_0^L \frac{\Pi(h)}{h + K_0}\, dx\right). \tag{6}$$

Additionally, following results from [50], model (3) has the energy functional

$$E[h] = \int_0^L \frac{1}{2}\left(\frac{\partial h}{\partial x}\right)^2 + U(h) \, dx, \tag{7}$$

where using (3), the rate of dissipation of energy can be obtained as

$$\frac{dE}{dt} = -\int_0^L h^3 \left(\frac{\partial p}{\partial x}\right)^2 dx + \gamma \int_0^L \frac{p^2}{h+K_0} \, dx. \tag{8}$$

The sign coefficient for the evaporation term, $\gamma$, will be seen to play an important role in determining the qualitative behavior of the model.

For $\gamma = 0$, model (3) reduces to (1), where mass is conserved and the energy is monotone decreasing. Bertozzi et al. [11] proved the global existence and positivity of solutions to the one-dimensional model (1) with nice initial data. They used the disjoining pressure defined by $\Pi(h) = Ah^{-3}(1 - \epsilon/h)$ to characterize the balance between the attractive and repulsive forces at film layer $h = O(\epsilon)$. This form of $\Pi(h)$ is identical to our generalized disjoining pressure (3d) with $P_0 = 0$, and the resulting potential is bounded from below. Their theorem can be applied to (3) with $\gamma < 0$ and $P_0 \geq 0$ [32] and is rephrased below.

**Theorem** (Bertozzi et al. [11]). *If potential $U(h)$ has a lower bound and initial data satisfies $h_0 > 0$, $h_0 \in H^1(\Omega)$ and $E[h_0] < \infty$, then a unique positive smooth solution for problem (1) exists for all $t > 0$.*

In their classical Schauder approach, the a priori lower bound of $U(h)$ together with energy dissipation provide a positive lower bound for the solution.

For $\gamma < 0$, equation (8) shows that the energy (7) with $\Pi(h)$ in (3d) is monotonically decreasing in time; for $P_0 \geq 0$ the potential $U(h)$ in (4) has a lower bound and the proof in the Theorem from [11] can be directly applied to this case.

In fact, with $\gamma \leq 0$ the requirement for $U(h)$ to have a lower bound is not necessary. With $P_0 < 0$, $\Pi(h)$ in (3d) leads to $U(h)$ with no lower bound. However, with $\gamma = 0$, the mass conservation and energy dissipation of model (1) still guarantee the positivity of the solution. While for $\gamma < 0$, a more sophisticated argument is needed to show the boundedness of the solution from below for model (3) [32].

Note that for $\gamma \neq 0$, where the conservation of mass in model (3) is not guaranteed, the change in total mass depends on both $\Pi(h)$ and $\gamma$. If the sign of $\Pi(h)$ varies with respect to the film thickness $h$, there is a rich family of equilibria with various dynamical behaviors for the model for different ranges of $\gamma$ and $P_0$. To simplify the discussion and fix the sign of $dM/dt$, we only consider the case with $\Pi(h) < 0$ for all film thickness $h$ and set $P_0 = -1$ for the rest of the paper. A full discussion of the range of behaviors in the evaporation/condensation model in different parameter regimes will be given in [32].

When $\gamma$ is positive more complicated behavior can occur, as suggested by the fact that the energy is no longer guaranteed to be monotone decreasing (8). With $\Pi(h) < 0$ equation (6) gives $dM/dt < 0$, which indicates that the dynamics of any film profile involves purely evaporative effects. Although (6) does not force $h_{\min}(t)$ to monotonically decrease in time, it is suggestive of either filmwise thinning or pointwise rupture driven by evaporation. Specifically for strong evaporation with $\gamma = 1$ numerical evidence in Figure 1(right) exhibits finite time rupture due to instabilities enhanced by the evaporation effects. This motivates us to investigate different instabilities of the model (3).

Also interestingly, when only weak evaporation is taken into consideration both dewetting phenomenon and finite-time rupture can be observed. A typical dynamical evolution with weak evaporation effects ($\gamma = 0.001$) is shown in Figure 3. While the early stage development of dry spot is similar to the nucleation observed in the no evaporation case ($\gamma = 0$) in Figure 1 (left), the evaporation effects cause slow droplet shrinkage followed by rupture in the precursor layer as the critical time $t_c \approx 3315.5$ is approached. Compared to the finite-time singularity induced by strong evaporation effects with $\gamma = 1$ in Figure 1 (right) where it takes only $t_c \approx 8.33$ to form the rupture starting from the same initial conditions, weak evaporation leads to much slower phases of dewetting evolution until the disjoining pressure is suppressed by evaporation in the later stage.

While the derivation of the evaporative flux (3b) given by Ajaev [2] suggests that the physically achievable range of parameters for typical fluids has $\gamma < 0$, in the spirit of other applied studies of singularity formation in physical systems, we will consider the rich and novel dynamics that occur for $\gamma > 0$ to better understand the influence of non-conservative effects and singular potentials on the development of finite-time rupture. This is also relevant to nonlinear interfacial instabilities in other fourth-order nonlinear PDE's with non-conservative contributions [3, 37, 35].

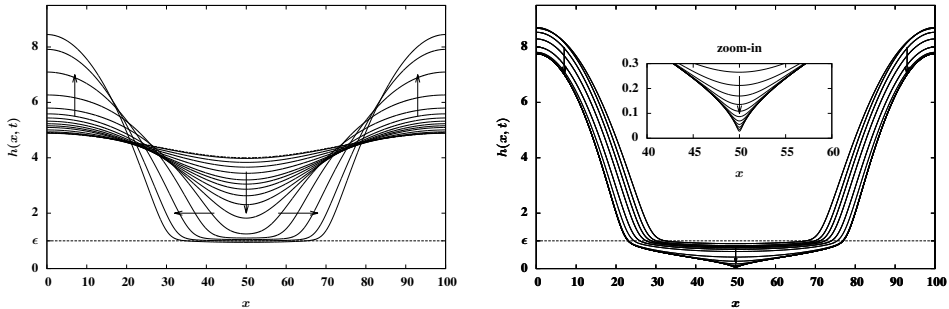

Figure 3: Evolution with weak evaporation, $\gamma = 0.001$, in model (3) with identical initial conditions and system parameters as in Figure 1. Early stage dewetting phenomenon (left) is followed by evaporation-dominated behavior with finite-time rupture in the later stage (right).

For the purpose of studying the rupture behaviors induced by intermolecular forces and evaporation effects, strong evaporation effects with $\gamma \equiv 1$ are assumed for the rest of the paper.

Singularity formation through self-similar dynamics has been studied extensively in many classes of PDE models, see for example [18, 52, 10]. In several problems involving fourth-order PDEs, it was seen that the regularizing influence of the highest order terms play a role in determining the form of the similarity solutions [37, 7, 53, 8]. In section 2 we will show rupture in (3) proceeds through two stages, each of which described by a self-similar solution solving a second order problem. These observations motivate us to explore how the form of our thin film model can lead to different types of rupture behaviors. In particular, guided by the overview given by Thiele [50], in section 3 we will generalize (3) to

$$\frac{\partial h}{\partial t} = \frac{\partial}{\partial x}\left(h^n \frac{\partial \tilde{p}}{\partial x}\right) + \frac{\tilde{p}}{h^m} \qquad \text{where} \qquad \tilde{p} = -\left(\frac{1}{h^4} + \frac{\partial^2 h}{\partial x^2}\right), \qquad (9)$$

which will allow us to focus attention on how conservative and non-conservative (here, evaporative) fluxes interact to influence the dynamics. In (9), the $(n, m)$ are the exponents in the mobility coefficients of the conserved and evaporative fluxes respectively, and this model has direct generalizations of the key properties of model (3), namely (5–8). We will show that (9) exhibits several different classes of rupture dynamics depending on $(n, m)$.

## 2. Self-similar rupture in model (3)

Here we will show that the dynamics leading to rupture at $(x_c, t_c)$, for $t \to t_c$ the solutions of model (3) can be expressed in terms of self-similar solutions of the form

$$h(x, t) \sim \tau^\alpha H(\eta), \qquad \tau = t_c - t, \qquad \eta = \frac{x - x_c}{\tau^\beta}, \qquad (10)$$

with $\alpha, \beta > 0$. The scaling parameter $\alpha > 0$ describes finite-time rupture occurring at $t = t_c$ with $h \to 0$ and $\beta > 0$ corresponds to spatial focusing at $x_c$ as $\tau \to 0$. Further, in localized rupture behavior while a singularity is approached at point $x_c$, the far-field must match to the regular behavior of the slowly evolving solution away from $x_c$ as $t \to t_c$ [8, 53]. This corresponds to a bounded time derivative, $h_t$, at any fixed point away from $x_c$ and yields the far-field boundary condition on the similarity solution,

$$\alpha H - \beta \eta H_\eta = 0 \quad \text{as } |\eta| \to \infty. \qquad (11)$$

We begin by combining the elements of model (3a-d) to write the governing evolution equation as

$$\frac{\partial h}{\partial t} = \frac{\partial}{\partial x}\left(h^3 \frac{\partial}{\partial x}\left[\Pi(h) - \frac{\partial^2 h}{\partial x^2}\right]\right) + \frac{1}{h+K_0}\left[\Pi(h) - \frac{\partial^2 h}{\partial x^2}\right] \tag{12}$$

where $\Pi(h) = h^{-3} - h^{-4} - 1$. Noting from (6) that the mean height will be monotone decreasing, in numerical simulations of (12) we have observed that when rupture occurs, the minimum film thickness is also decreasing, with the solution taking the form (10) in a neighborhood of the minimum.

Since rupture will not occur without the evaporative term, we expect it to play a key role in determining the dynamics. Consequently, based on the form of that term, we also expect the dynamics to change according to the relationship between $K_0$ and $h_{\min}(t)$. We divide the evolution into two stages: the early stage with $h_{\min}(t) \gg K_0$ and the later stage with $h_{\min}(t) \ll K_0$.

2.1. Early stage dynamics for $h_{\min}(t) \gg K_0$

For initial conditions with $h_{\min} \gg K_0$, in the vicinity of $x_c$ the evaporation term can be approximated by

$$\frac{p(h)}{h+K_0} \sim \frac{p(h)}{h}\left(1 + \frac{K_0}{h} + \cdots\right) \sim \frac{p(h)}{h}, \tag{13}$$

then the leading order form of (12) after we substitute-in (10) is

$$\tau^{\alpha-1}\left(-\alpha H + \beta\eta H_\eta\right) = \left(-3\tau^{-2\beta}\left(\frac{H_\eta}{H}\right)_\eta + 4\tau^{-\alpha-2\beta}\left(\frac{H_\eta}{H^2}\right)_\eta - \tau^{4\alpha-4\beta}\left(H^3 H_{\eta\eta\eta}\right)_\eta\right) \\ + \left(\tau^{-4\alpha}\frac{1}{H^4} - \tau^{-5\alpha}\frac{1}{H^5} - \tau^{-2\beta}\frac{H_{\eta\eta}}{H} - \tau^{-\alpha}\frac{1}{H}\right), \tag{14}$$

where the terms are grouped according to the rate of change (on the left) and the conservative and evaporative fluxes on the right.

Exact similarity solutions exist when a PDE has an invariant scaling symmetry, such that a choice of $\alpha, \beta$ in (10) can be found to separate out $\tau$ to reduce the PDE to an ordinary differential equation for the similarity profile $H(\eta)$. Due to the number of different terms present in (14), this is not possible, but we can seek an asymptotically self-similar solution, namely the scale-invariant solution of the leading order dominant balance of terms for the limit $\tau \to 0$.

For $\tau \to 0$ with $H = O(1)$, there are four possible leading-order terms in (14),

$$\tau^{\alpha-1}\left(-\alpha H + \beta\eta H_\eta\right), \qquad 4\tau^{-\alpha-2\beta}\left(\frac{H_\eta}{H^2}\right)_\eta, \qquad -\tau^{4\alpha-4\beta}\left(H^3 H_{\eta\eta\eta}\right)_\eta, \qquad -\tau^{-5\alpha}\frac{1}{H^5}.$$

With these, there are only two feasible dominant balances for dynamic solutions,

$$\begin{cases} (a) & \text{If } \alpha - 1 = -\alpha - 2\beta = 4\alpha - 4\beta \text{ then } \alpha = \frac{1}{7}, \beta = \frac{5}{14}, \\ (b) & \text{If } \alpha - 1 = -5\alpha = -\alpha - 2\beta \text{ then } \alpha = \frac{1}{6}, \beta = \frac{1}{3}. \end{cases} \tag{15}$$

If the first scaling relationship is adopted, then (10) yields

$$h(x,t) = (t_c - t)^{1/7} H\left(\frac{x - x_c}{(t_c - t)^{5/14}}\right),$$

where $H(\eta)$ satisfies the fourth-order ODE,

$$\frac{1}{7}\left(H - \frac{5}{2}\eta H'\right) + 4\left(\frac{H'}{H^2}\right)_\eta - \left(H^3 H'''\right)_\eta = 0,$$

corresponding to the leading order PDE

$$h_t = -(h^3 h_{xxx})_x + 4(h^{-2} h_x)_x. \tag{16}$$

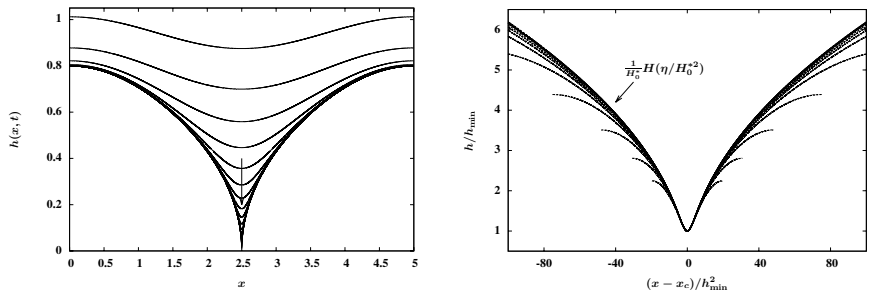

Figure 4: (Left) Numerical solution of (12) with $K_0 = 0$ and initial data $h_0(x) = 1.2 + 0.1\cos(2\pi x/5)$ leading to finite-time rupture; (Right) the numerical profiles appropriately rescaled by $h_{\min}(t)$ converging to the similarity solution (17) scaled as $H(\eta/(H_0^*)^2)/H_0^*$ of the leading order ODE (18).

This is a thin-film type equation with stabilizing diffusion [40], where both nonlinear terms work to smooth the solution and prevent rupture. The positivity of solutions for all $t > 0$ can be demonstrated by means of the Theorem from [11], applied with $U$ given by $U(h) = \frac{1}{3}h^{-3}$. Consequently, this choice of scaling cannot yield rupture.

In contrast, the second choice of scalings, (15b), yields

$$h(x,t) = (t_c - t)^{1/6} H\left(\frac{x - x_c}{(t_c - t)^{1/3}}\right), \tag{17}$$

where $H(\eta)$ solves the second-order ODE,

$$\frac{1}{6}(H - 2\eta H_\eta) - \frac{1}{H^5} + 4\left(\frac{H_\eta}{H^2}\right)_\eta = 0, \tag{18}$$

and (11) reduces to the boundary condition $H - 2\eta H_\eta = 0$ as $|\eta| \to \infty$. A one-parameter shooting method starting from $H(0) = H_0$ and $H'(0) = 0$ is used to numerically solve the similarity equation (18) subject to (11). The numerical results suggest that there is a unique self-similar solution $H(\eta)$ with $H(0) = H_0^* \approx 1.2934$. Figure 4 depicts a comparison between the dynamics of PDE (12) with $K_0 = 0$ and the similarity solution of (18). This confirms that the scaled PDE solution converges to $H(\eta)$ as $t \to t_c$.

The nonlinear PDE (12) was solved numerically using a fully implicit second-order finite difference method with both a uniformly-spaced fixed grid and an adaptive non-uniform grid. In particular, we used the Keller box method [33] applied to the fourth-order equation expressed as a discretized, cell-centered system of four first-order differential equations for $h, p, q = p_x$ and $s = h_x$. To better capture the rapidly changing finite-time rupture with high resolution, we use a classical moving mesh scheme in quadruple precision with a tailored monitor function to assign clustered grid points around singularity position. Detailed discussions and applications of related adaptive mesh methods can be found in [14, 37].

A simple way to quantitatively access if the PDE dynamics coincide with the analytical predictions is obtained by noting that (17) yields that $h(x_c, t) = (t_c - t)^{1/6} H(0)$ and $h_{xx}(x_c, t) = (t_c - t)^{-1/2} H''(0)$. Consequently, as rupture is approached we expect that the PDE solution at $h_{\min}$ satisfies,

$$h_{xx}(x_c, t) \sim C^* h(x_c, t)^{-3} \qquad \text{as } t \to t_c, \tag{19}$$

and using (18), the scaling constant is given by $C^* = (1 - H_0^{*6}/6)/4 \approx 0.055$. Figure 5 confirms the relationship (19) and illustrates the numerical saturation effects observable in finite-resolution simulations in fixed-grid finite-difference simulations compared with the adaptive non-uniform grid scheme that will be used for all further simulations. Since (19) can be evaluated without the need to estimate $t_c$, we will use relations of this form to identify the nature of the rupture dynamics in further cases to be studied.

The PDE corresponding to (18) is

$$h_t = 4\left(h^{-2} h_x\right)_x - h^{-5}. \tag{20}$$

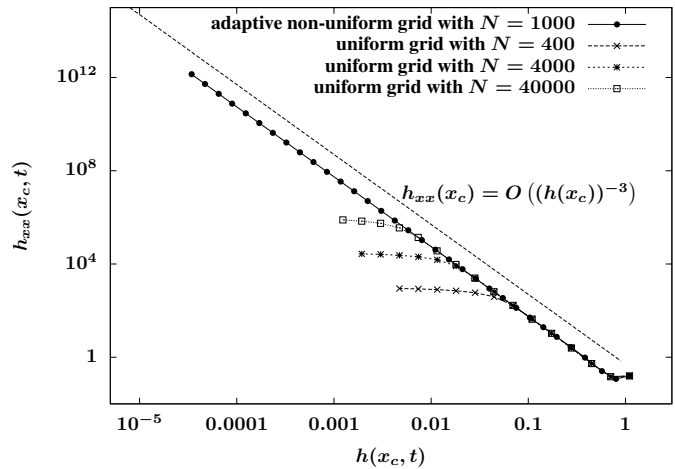

Figure 5: Plot of $h_{xx}(x_c, t)$ vs $h(x_c, t)$ for numerical simulations of Figure 4 with increasing spatial resolution in uniform grids and with $N = 1,000$ grid points in an adaptive computational grid, showing convergence and agreement with prediction (19).

There is vast literature on singularity formation in second-order reaction-diffusion, for example see [36, 27]. Equation (20) belongs to the family of quasilinear diffusion equations with absorption [5, 46]

$$u_t = (u^{\sigma-1} u_x)_x - u^\rho \qquad (21)$$

which are further classified for different range of parameters $\sigma$ and $\rho$. In particular, (20) is categorized as a fast diffusion equation ($\sigma < 1$) with singular absorption ($\rho < 1$). From perspectives of existence, uniqueness and regularity, Ferreira and Vazquez [20] studied the finite-time extinction behavior of solutions to the Cauchy problem of (21) with the range of parameter $0 < (\sigma, \rho) < 1$ and compactly supported initial data. In many other contexts, singular behavior for $u \to 0$ is called quenching [36]. There have been many studies of self-similar and non-self-similar solutions for finite-time quenching in general quasilinear and semilinear ($\sigma = 1$) forms of (21) [21, 22, 23]. In one-dimension, a critical value for the absorption, $\rho^* = 2 + \sigma$ marks a transition between different forms of dynamics [27]. In applications to micro-electro-mechanical systems (MEMS), finite-time quenching describing "touch-down" or "pull-in" of a flexible capacitor were studied, most typically with $\sigma = 1$ and $\rho = -2$ [19, 37, 24].

2.2. Later stage dynamics for $h_{\min}(t) \ll K_0$

For $h_{\min}(t) \ll K_0$, in a neighborhood of $x_c$ the evaporative term can be expanded in terms of $h/K_0$ with

$$\frac{p(h)}{h + K_0} \sim \frac{p(h)}{K_0}\left(1 + \frac{h}{K_0} + \cdots\right) \sim \frac{p(h)}{K_0}, \qquad (22)$$

and the leading order equation for (12) as $h/K_0 \to 0$ then takes the form

$$h_t = (h^3 p_x)_x + \frac{p}{K_0}. \qquad (23)$$

We then use the scalings (10) to seek asymptotic similarity solutions to (23). Similar analysis as in section 2.1 leads to the only feasible scaling constants as $\alpha = 1/5$ and $\beta = 3/10$ yielding the similarity solution

$$h(x,t) = (t_c - t)^{1/5} H\left(\frac{x - x_c}{(t_c - t)^{3/10}}\right), \qquad (24)$$

where $H(\eta)$ satisfies the ODE,

$$\frac{1}{5} H - \frac{3}{10}\eta H' - \frac{1}{K_0 H^4} + 4\left(\frac{H'}{H^2}\right)_\eta = 0. \qquad (25)$$

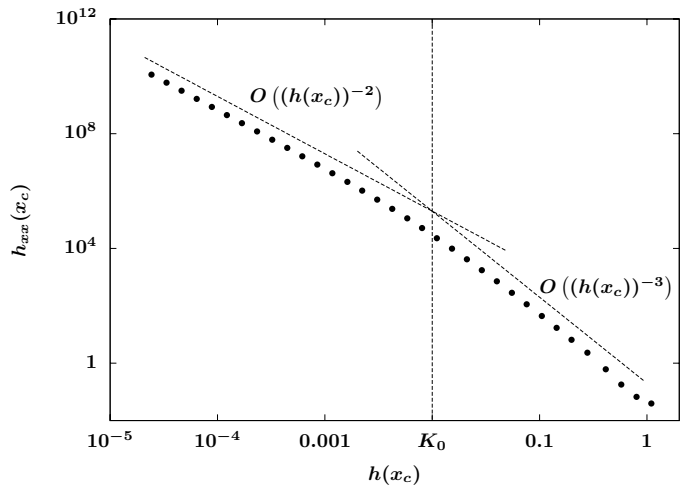

Figure 6: Plot of $h_{xx}(t, x_c)$ vs $h(t, x_c)$ for (12), as in Figure 4 except with $K_0 = 0.01$. The transition between the early-stage (19) and the later-stage (27) self-similar behavior is visible as $h_{\min}(t) = h(x_c, t)$ crosses below $K_0$.

The corresponding leading order PDE for this case is

$$h_t = 4\left(h^{-2}h_x\right)_x - \frac{1}{K_0 h^4}. \tag{26}$$

Solution (24) yields that $h(x_c, t) = (t_c - t)^{1/5} H(0)$ and $h_{xx}(x_c, t) = (t_c - t)^{-2/5} H''(0)$, hence like (19), here we obtain

$$h_{xx}(x_c, t) \sim C h(x_c, t)^{-2} \qquad \text{as } t \to t_c. \tag{27}$$

Figure 6 shows the transition between the early-stage and later-stage stage similarity solutions as rupture is approached, with $h_{\min} \to 0$ in (12) with a finite value of $K_0$.

### 3. Rupture behavior in a non-conservative generalized thin-film equation

It is interesting to note that the regularizing fourth order term due to surface tension in (12) does not play a role in determining the form of the evaporation-dominated rupture dynamics. We wish to put that result in context by embedding that equation in a generalized model that will allow us to study how the competition between evaporation and dewetting can lead to finite-time rupture.

In this section we will examine dynamics leading to rupture in the generalized lubrication equation (9) parametrized by the exponents in the mobility coefficients $(n, m)$ in the conservative and evaporative terms respectively,

$$\frac{\partial h}{\partial t} = -\frac{\partial}{\partial x}\left(h^n \frac{\partial}{\partial x}\left[\frac{1}{h^4} + \frac{\partial^2 h}{\partial x^2}\right]\right) - \frac{1}{h^m}\left[\frac{1}{h^4} + \frac{\partial^2 h}{\partial x^2}\right], \tag{28}$$

subject to the same Neumann boundary conditions (3e) as in model (3) and (12). We note that the pressure has been reduced to the dominant terms for the limit $h \to 0$ using $h^{-4} \gg h^{-3} \gg P_0$. Model (12) with $K_0 = 0$ is a special case of (28) for $n = 3$ and $m = 1$, and the late-stage case, equation (23), corresponds to (28) with $n = 3$ and $m = 0$.

Our generalized model is partly motivated by the study by Bertozzi and Pugh [9], where conditions for global existence of solutions versus finite-time blow-up with respect to the exponents of the mobility coefficients of competing second- and fourth-order terms in

$$u_t = -(u^b u_{xxx})_x - (u^a u_x)_x,$$

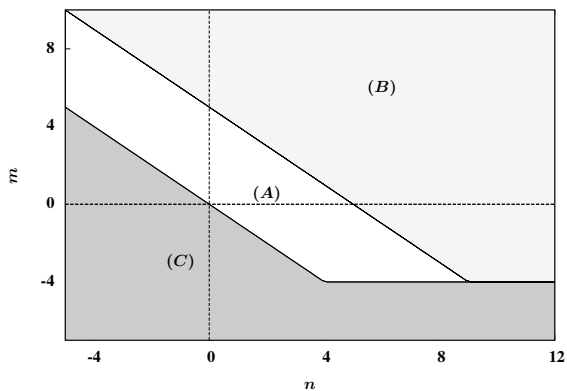

Figure 7: A preliminary version of the bifurcation diagram for the dynamics of (28) with respect to exponents $(m, n)$ that separates dynamics given by linear stability analysis (region $(C)$) from two classes of finite-time rupture solutions, regions $(A, B)$.

were determined. Further properties of solutions can be found in [40, 17]. Some results for the extension of this model with an additional nonconservative term,

$$u_t = -(u^b u_{xxx})_x - (u^a u_x)_x - u,$$

were given in [35]. With $b = 0$ this class of PDE was also studied by [3] and in the context of solidification, sometimes called the Sivashinsky equation, in [47]. Without the second-order term, some classes of solutions for PDEs of the form

$$u_t = -(u^b u_{xxx})_x - u^\rho$$

have been studied by [25]. Our study of rupture has closest connections to the work on finite-time touch-down solutions for MEMS applications for this PDE with $b = 0$ and $\rho = -2$ [37].

The behavior of solutions of (28) depend on the values of the parameters $m$ and $n$ that control the competing influences of destabilizing evaporative loss (non-conservative effects) and regularizing conservative surface tension effects. In particular, note that the asymptotic reduction of the pressure yields a pure wetting potential, $\Pi \sim -h^{-4}$, as in equation (16), for which rupture cannot occur in the absence of the evaporative term. Hence rupture must always be driven by the nonconservative term, but we will show that depending on $(n, m)$ the dynamics can take very different forms.

Our goal is to construct a bifurcation diagram indicating the behavior at each $(n, m)$. Figure 7 gives a preliminary form of this, based on only two considerations: self-similar rupture solutions (as in §2) and linear stability of flat solutions. In section 3.1 we will show that in one regime (region $(C)$), point rupture does not occur, but linear stability indicates that spatial variations decay out to lead to uniform thinning in the solution.

Outside $(C)$, instabilities cause perturbations to grow and we show that rupture can occur in finite time. Assuming the solution to be a self-similar solution of form (10), $h(x, t) = \tau^\alpha H(\eta)$, the generalized model (28) reduces to

$$\tau^{\alpha-1}\left(-\alpha H + \beta\eta H_\eta\right) = -\left(-4\tau^{(n-4)\alpha-2\beta}\left(H^{n-5}H_\eta\right)_\eta + \tau^{(n+1)\alpha-4\beta}\left(H^n H_{\eta\eta\eta}\right)_\eta\right) \\ - \left(\tau^{-(4+m)\alpha}\frac{1}{H^{4+m}} + \tau^{(1-m)\alpha-2\beta}\frac{H_{\eta\eta}}{H^m}\right). \tag{29}$$

As in (14), the dominant balance of terms in (29) as $\tau \to 0$ determines the equation satisfied by the asymptotic similarity solution. It can be shown that by comparing the exponents of the two terms,

$$-4\tau^{(n-4)\alpha-2\beta}\left(H^{n-5}H_\eta\right)_\eta \quad \text{and} \quad \tau^{(1-m)\alpha-2\beta}\frac{H_{\eta\eta}}{H^m},$$

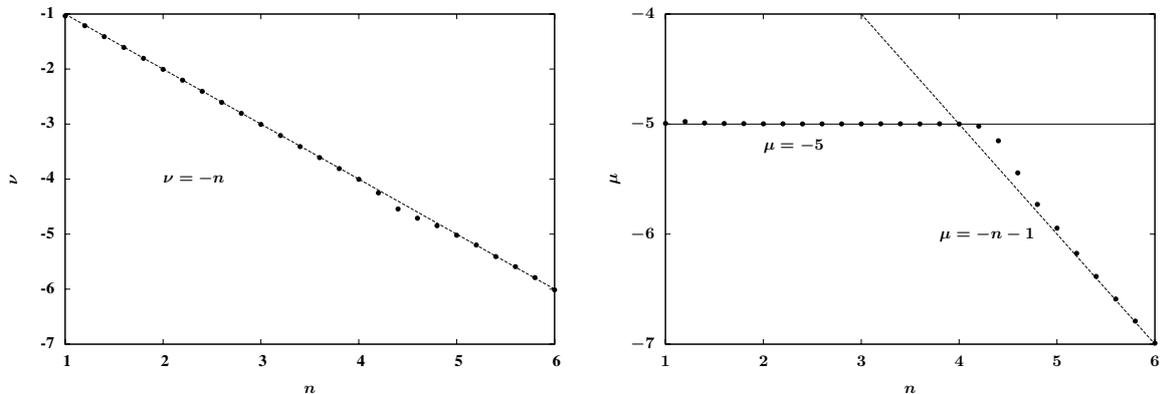

Figure 8: (Left) The value of $\nu$ in (31) obtained by a linear regression of $\log(h(x_c,t))$ and $\log(h_{xx}(x_c,t))$ from simulations of (28) over a range of $n$. (Right) Similarly, $\mu$ in (32) estimated from the same simulations from linear regression of $\log(h(x_c,t))$ and $\log|h_t(x_c,t)|$. All simulations were done with $m=1$ and domain-size $L=1$ starting from initial conditions $h_0(x) = 0.5 - 0.01\exp(-100(x-1/2)^2)$.

the form of the dominant balance depends on the sign of $n+m-5$. Consequently, exterior to $(C)$, we can identify two regions, called $(A)$ and $(B)$, corresponding to the cases $n+m-5 < 0$ and $n+m-5 > 0$ respectively. In sections 3.2 and section 3.3 we will show that these regions yield fundamentally different classes of finite-time rupture solutions (and further analysis will uncover additional distinctions).

Applying the assumed form of the self-similar solution (10) at the position of the rupture yields that

$$h(x_c,t) = \tau^\alpha H(0), \quad h_{xx}(x_c,t) = \tau^{\alpha-2\beta} H''(0), \quad h_t(x_c,t) = -\alpha\tau^{\alpha-1} H(0). \tag{30}$$

Eliminating $\tau$ between the first two expressions, we obtain the relation

$$h_{xx}(x_c,t) = C h(x_c,t)^\nu \qquad \text{where} \quad \nu = 1 - \frac{2\beta}{\alpha}, \tag{31}$$

where the coefficient is uniquely determined by the local properties of $H(\eta)$; this relation generalizes (19, 27) used earlier. Similarly, using the first and third expressions from (30) yields

$$|h_t(x_c,t)| = \tilde{C} h(x_c,t)^\mu \qquad \text{where} \quad \mu = 1 - \frac{1}{\alpha}. \tag{32}$$

Figure 8 shows results from a series of numerical simulations of (28) with $m=1$ and a range of $n$ values spanning regions $(A,B)$. Finite-time rupture was observed in each and linear regression was used to estimate $\nu,\mu$ from the simulation data. The figure suggests that a single relation for $\nu$ holds over the whole range of $n$, but that $\mu$ shows a bifurcation in behavior at $n=4$, where $n+m-5$ changes sign, in-line with our expectations from (29).

3.1. Region $(C)$: Uniform thinning

Finite-time singularities in our model can result from growth in spatial variations due to strong instabilities in smooth solutions. However, for some parameter ranges, spatially uniform solutions to model (28) with small thickness are actually stable and thus cannot lead to rupture. In this section we investigate the evolution of these uniform solutions to model (28) via linear stability analysis. We will demonstrate that for parameters $(n,m)$ falling in Region $(C)$, which is defined by

$$(C) = \{(n,m): \ m+n < 0 \text{ or } m < -4\}, \tag{33}$$

flat solutions $h = \bar{h}(t)$ to model (28) with $\bar{h} \ll 1$ are linearly stable.

To study the stability of a flat solution, we perturb it by individual Fourier mode disturbances

$$h(x,t) = \bar{h}(t) + \delta e^{ikx} e^{\sigma(t)} + O(\delta^2), \tag{34}$$

where $k$ is the wavenumber, $\sigma(t)$ represents the growth of disturbances in time starting from the initial amplitude $\delta \ll 1$. Substituting (34) into model (28) and linearizing about $h = \bar{h}$ yield the $O(1)$ and $O(\delta)$ equations:

$$O(1): \quad \frac{d\bar{h}}{dt} = -\bar{h}^{-(4+m)}, \tag{35a}$$

$$O(\delta): \quad \frac{d\sigma}{dt} = \left(k^2 \bar{h}^{-m} + (m+4)\bar{h}^{-(m+5)}\right) - \left(k^4 \bar{h}^n + 4k^2 \bar{h}^{n-5}\right). \tag{35b}$$

The $O(1)$ equation (35a) indicates that starting from $\bar{h}(0) = \bar{h}_0$, the leading order behavior of $\bar{h}(t)$ follows

$$\bar{h}(t) = \begin{cases} \left(\bar{h}_0^{5+m} - (5+m)t\right)^{1/(5+m)} & m \neq -5 \\ \bar{h}_0 e^{-t} & m = -5 \end{cases}. \tag{36}$$

As is expected from (6), $\bar{h}$ is decreasing in time for all values of $m$ due to the evaporative term. For $m > -5$ we expect the solution to vanish in finite time and uniform thinning, $\bar{h}(t) \to 0$, occurs as the critical time $t_c = (5+m)^{-1} \bar{h}_0^{5+m}$ is approached; while for $m \leq -5$ infinite-time thinning occurs. In particular, for $m = -5$, the thin film thickness decays exponentially. This qualitative behavior, $h(x, t_c) \equiv 0$ for some $t_c > 0$, is sometimes called "total extinction" or "complete extinction", and has been studied for a wide range of fast diffusion equations [20]. Based on whether the critical extinction time $t_c$ is finite or not, we subdivide Region $(C)$ into two parts:

- Region $(C_1)$: finite-time uniform thinning with $m > -5$.
- Region $(C_2)$: infinite-time uniform thinning with $m \leq -5$.

Note that localized rupture can be expected to occur from a perturbed flat solution only if the solution is unstable with small film thickness $\bar{h}$. On the right-hand side of the $O(\delta)$ equation (35b), the last two terms are always stable for any system parameters, while the first term is always destabilizing, and the second term $(m+4)\bar{h}^{-(m+5)}$ is stable only if $m < -4$. To determine the sign of $d\sigma/dt$ for $\bar{h} \ll 1$, we compare the exponents of the stable and unstable terms, and conclude that the stable terms overcome the unstable ones if $m < -4$ or $m+n < 0$. Therefore, for $\bar{h} \ll 1$ we have the dependence of dispersion relation on parameters $(n, m)$,

$$\begin{cases} d\sigma/dt < 0 & \text{if } m < -4 \text{ or } m+n < 0, \\ d\sigma/dt > 0 & \text{if } m > -4 \text{ and } m+n > 0. \end{cases} \tag{37}$$

Here a positive linear growth rate $d\sigma/dt$ of disturbances for small film thickness suggests the possibility of rupture phenomenon. Therefore the relationship (37) indicates that the parameters $(n, m)$ in region $(C)$ do not allow finite-time rupture phenomenon for near-flat small initial data. Instead the spatial perturbation decays quickly in time and uniform thinning will occur as $\bar{h} \to 0$.

To get further information on the evolution of spatial disturbances, we rewrite $\sigma$ in terms of $\bar{h}$ using both equations in (35),

$$\sigma(\bar{h}(t)) = \begin{cases} -(m+4)\ln(\bar{h}) - \frac{1}{5}k^2 \bar{h}^5 + \frac{4}{m+n}k^2 \bar{h}^{m+n} + \frac{k^4}{m+n+5}\bar{h}^{n+m+5} + a_1, & m+n \neq -5, 0, \\ (4k^2 - m - 4)\ln(\bar{h}) + \frac{1}{5}(k^4 - k^2)\bar{h}^5 + a_2. & m+n = 0, \\ -\frac{4k^2}{5}\bar{h}^{-5} + (k^4 - m - 4)\ln(\bar{h}) - \frac{k^2}{5}\bar{h}^5 + a_3 & m+n = -5, \end{cases} \tag{38}$$

where $a_1, a_2, a_3$ are constants depending on initial conditions. Noting that the form (34) yields $h_{xx}(x_c, t) \sim -\delta k^2 \exp(\sigma(t))$, we derive from (38) that for perturbed flat solutions with $\bar{h} \ll 1$, the leading order relationship between the linearized curvature at the critical position $x_c$, $h_{xx}(x_c, t)$, and $\bar{h}$ is

$$h_{xx}(x_c, t) \sim c_1 \bar{h}^{-(m+4)}, \qquad m+n > 0, \tag{39a}$$

$$h_{xx}(x_c, t) \sim c_2 \bar{h}^{4k^2 - (m+4)}, \qquad m+n = 0, \tag{39b}$$

$$h_{xx}(x_c, t) \sim c_3 \exp\left(\bar{h}^{m+n}\right) \bar{h}^{-(m+4)}, \qquad m+n < 0, \tag{39c}$$

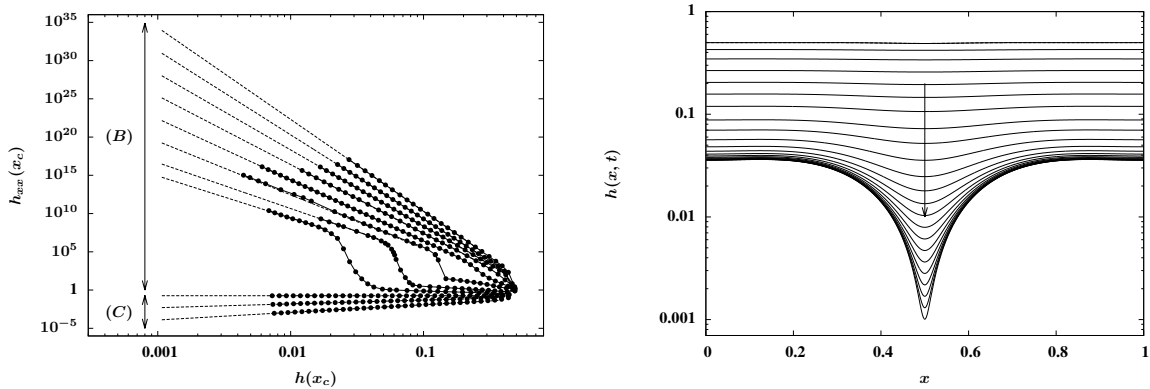

Figure 9: (Left) Plot of $h_{xx}(x_c)$ as a function of $h(x_c)$ with fixed $n = 10$ and over a range of $m$ values, $-5 \leq m \leq 3$, with $m = -5, -4.5, -4, -3.5, -3, -2, \cdots, 3$. The dividing line between $(B)$ and $(C)$ is $m = -4$. In region $(C)$, uniform thinning occurs all the way to $h \to 0$, otherwise (here corresponding to region $(B)$), loss of stability will eventually yield localized rupture. Here dots represent direct PDE simulation results, and dashed lines are predictions of dynamics in (39a) for region $(C)$ and (31) with $\nu$ given by (73) for region $(B)$. (Right) A typical numerical simulation in region $(B)$, starting from identical initial data used in Figure 8 with system parameters $n = 10, m = -3.5$, showing early uniform thinning transitioning to finite-time rupture.

where $c_1, c_2, c_3$ are constants that depend on $k$, $m$ and $n$.

Therefore the sign of $m + n$ further separates the uniform thinning regime $(C)$ into two subcases. For $m + n < 0$, $h_{xx}(x_c)$ decays exponentially with respect to $\bar{h}$ (39c) as $\bar{h} \to 0$. While for $m + n > 0$, $h_{xx}(x_c)$ is asymptotically a power of $\bar{h}$ (39a) as $\bar{h} \to 0$, which is similar to the scaling form of the self-similar rupture solution (31). In particular, for $m + n > 0$ and $m < -4$, (39a) indicates that the spatial variation decays as $\bar{h}$ approaches zero, which corresponds to uniform thinning. Moreover, for $(n, m)$ satisfying $m + n > 0$ and $m > -4$, which is in the finite-time rupture regime, the relationship in (39a) provides a good prediction for the early-stage slowly-varying dynamics of solutions to (28) before localized singularity profiles are approached in the later stage.

A numerical verification of the relationships in (37) and (39a) is plotted in Figure 9 (left). For fixed $n = 10$ and varying $m$, as $h(x_c) \to 0$ the behavior of $h_{xx}(x_c, t)$ depends on the value of $m$. For $m < -4$ (uniform thinning regime), $h_{xx}(x_c) \to 0$ as $h(x_c) \to 0$ indicating that the film is uniformly flattening over time. The value $m = -4$ serves as a threshold between the uniform thinning region $(C)$ and the finite-time rupture case, where $h_{xx}(x_c)$ approaches a positive constant as $h(x_c) \to 0$. The linear stability analysis is useful even outside of region $(C)$. For $m > -4$, where finite-time singularities are expected to develop, the solution evolves slowly with small spatial variation following (39a) in the early stage. In the later stage we observe $h_{xx}(x_c) \to \infty$, which suggests rupture phenomenon. The scaling form between $h_{xx}(x_c)$ and $h(x_c)$ in the later stage as $t \to t_c$ will be derived in section 3.3. A simple transition between the two stages for $(n, m) = (10, -3.5)$, which is outside of region $(C)$, is shown in Figure 9 (right) as the growth of perturbations in the near-flat initial data finally leads to localized rupture. In order to determine the range for which the above linear stability analysis applies, here we use the average film thickness to approximate $\bar{h}$. The evolution of $\bar{h}$ and spatial disturbance $\bar{h} - h_{\min}$ are plotted in Figure 10, with their early-stage behavior matching the expected decay rate of $\bar{h}$ in (36) and the growth rate of the disturbance in (38). The transition between the early slowly-varying phase and the later finite-time rupture phase may also involve complicated intermediate spatial oscillations and general nonlinear behaviors.

Based on the above linear stability analysis result, we expect rupture to occur for model (28) with parameters $(n, m)$ outside of region $(C)$. In the following subsections we will use dimensional analysis and a local expansion method to derive further constraints on $(n, m)$ for possible self-similar and non-self-similar rupture solutions.

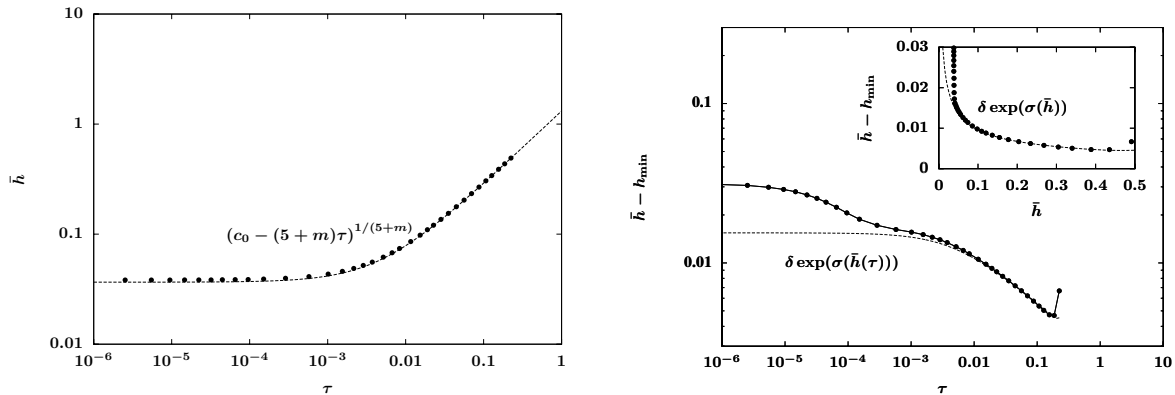

Figure 10: Plots of average film thickness $\bar{h}$ and the growth of disturbance $\bar{h} - h_{\min}$ for the numerical simulation in Figure 9 (right) with $(n, m) = (10, -3.5)$. Agreement with the linear stability results (36) and (38) for the early stage of evolution is clear.

*3.2. Region (A): Second order rupture*

From the similarity equation (29) of the generalized model (28), we start the analysis for self-similar finite-time rupture solutions with $(n, m)$ in parameter region $(A)$, where

$$(A) = \{(n, m) : 0 < m + n < 5, m > -4\}. \tag{40}$$

Recall that the early and later-stage self-similar rupture solutions for model (3) discussed in section 2 correspond to $(n, m) = (3, 1)$ and $(n, m) = (3, 0)$ in model (28), respectively. As both of the parameter pairs fall in Region $(A)$, we expect the self-similar rupture results for model (28) with parameters in this region to generalize the results in section 2.

For parameters $(n, m)$ in Region $(A)$, there are two possible dominant balances. But like the balance in section 2.1 the corresponding leading order differential equation for one of the dominant balance does not admit rupture behavior, therefore is not useful. The other feasible balance yields the leading order equation

$$O(\tau^{-\frac{m+4}{m+5}}): \quad -\alpha H + \beta \eta H_\eta + \frac{1}{H^{4+m}} - 4\left(H^{n-5} H_\eta\right)_\eta = 0, \tag{41}$$

with the scaling parameters

$$\alpha = \frac{1}{m+5}, \quad \beta = \frac{n+m}{2(m+5)}. \tag{42}$$

It follows from the scalings (42) and the relationships in (31, 32) that

$$\nu = 1 - n - m, \quad \mu = -m - 4. \tag{43}$$

Recall that in Figure 8, where $m = 1$, the observed scaling parameters $\nu = -n$ and $\mu = -5$ from numerical evidence are identical to the prediction in (43).

Apart from using linear stability analysis, a direct observation of the scaling parameters can also lead to the threshold between uniform thinning and second order rupture regime. Note that the scaling parameters in (42) need to satisfy $\alpha, \beta > 0$ for localized rupture solutions. This corresponds to $m + 5 > 0$ and $m + n > 0$. Also since $|h_t(x_c, t)|$ is expected to approach infinity as $h(x_c, t)$ approaches zero, $\mu < 0$ in (32), which corresponds to $m > -4$ in this case, is needed for localized rupture to happen in finite time.

Here we briefly discuss the behavior of the similarity solution $H(\eta)$ in the far field, as $|\eta|$ approaches infinity. Again the quasi-steady behavior of $h(x, t)$ away from $x_c$ leads to far-field boundary conditions (11). With the scaling (42) it reduces to

$$H - \frac{n+m}{2} \eta H_\eta = 0 \quad \text{as } |\eta| \to \infty, \tag{44}$$

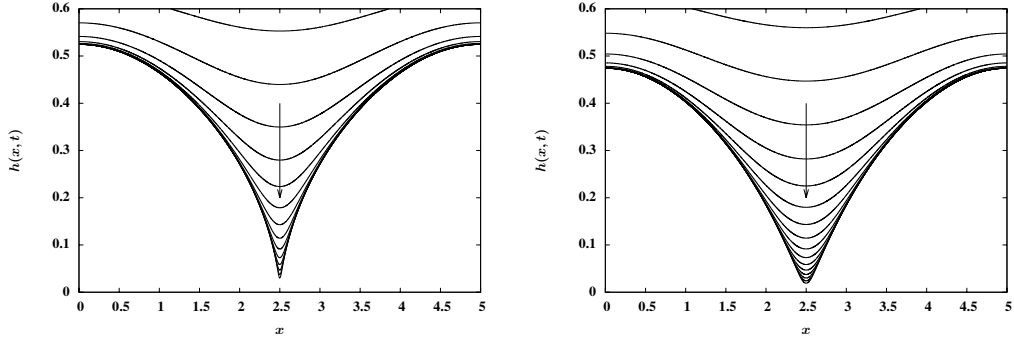

Figure 11: A comparison of rupture profiles with $m = 0$ and (left) $n = 3$ : case ($A_1$), (right) $n = 2$ : case ($A_2$), starting from identical initial conditions used in Figure 4.

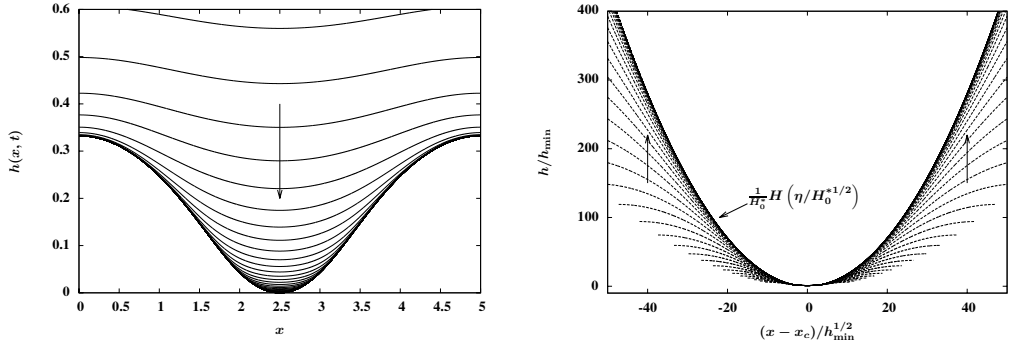

Figure 12: Second-order rupture in case ($A_3$) : $m = 0, n = 1$. (Left) Self-similar rupture solutions from direct PDE simulation with identical initial conditions used in Figure 4 and (right) rescaled PDE solutions showing convergence to the similarity solution scaled as $H(\eta/(H_0^*)^{1/2})/H_0^*$ of the leading order ODE (41), where $H_0^* = H(0) = 1.2019$.

which yields the asymptotic far-field behavior

$$H(\eta) \sim C\eta^{2/(n+m)} \quad \text{as } |\eta| \to \infty. \tag{45}$$

The form of (45) reveals that in addition to the usual 90° macroscopic contact angle rupture observed in Figure 4, rupture profiles with zero contact angles also exist. In particular, the dependence of the exponent in (45) on $m + n$ divides the second order self-similar rupture scenario into three subcases:

($A_1$) For $2 < n + m < 5$, $n < 5$ and $m > -4$, $H(\eta) \sim C|\eta|^k$ with $0 < k < 1$ as $|\eta| \to \infty$. In $(x,t)$ coordinates, the rupture has infinite slope and contact angle $\theta = 90°$, as is shown in Figure 11 (left).

($A_2$) For $n + m = 2$, $n < 5$ and $m > -4$, we have the borderline case $H(\eta) \sim C|\eta|$ as $|\eta| \to \infty$. In $(x,t)$ coordinates, the rupture has finite slope and contact angle $0° < \theta < 90°$, as in Figure 11 (right).

($A_3$) For $0 < n + m < 2$, $n < 5$ and $m > -4$, $H(\eta) \sim C|\eta|^k$ with $k > 1$ as $|\eta| \to \infty$. In $(x,t)$ coordinates, the rupture has zero slope and contact angle $\theta = 0°$, as in Figure 12 (left).

In Figure 12 we observe that $h_{xx}(x_c)$ approaches a constant as the rupture develops, which matches the prediction in (43) with $\nu = 0$. In fact, for $0 < m + n < 1$, localized rupture occurs, with vanishing local curvature, $h_{xx}(x_c) \to 0$ as $h(x_c) \to 0$, based on the scaling parameter $\nu$ given by (43). A detailed illustration of the relationship between $h_{xx}(x_c)$ and $h(x_c)$ for system parameters in different ranges will be shown later in Figure 14.

The corresponding leading order PDE for (41) in $(x,t)$ variables is given by

$$h_t = 4(h^{n-5}h_x)_x - \frac{1}{h^{m+4}}. \tag{46}$$

This equation belongs to the family of diffusion equations with absorption (21) with $\sigma = n - 4$ and $\rho = -(m+4)$, where $m > -4$ indicates that the absorption term is singular. In particular, the threshold for fast diffusion, $\sigma < 1$, which corresponds to $n < 5$, suggests the possibility of an additional bifurcation point on the system parameters $(n, m)$. We will show in section 3.2.1 that $n < 5$ is actually a necessary condition for the validity of self-similar rupture solutions (10).

*3.2.1. Refined analysis of the self-similar solution for (A)*

Following the method used in [30], we apply a local expansion method to a transformed version of (28) to derive the constraint $n < 5$ for the second-order self-similar rupture solution (10, 42). This new constraint will lead to a redefined region $(A)$.

Motivated by the leading order solution (36) from the linear stability result, we introduce the transformation
$$h(x,t) = \left((m+5)v(x,\tau)\right)^{1/(m+5)}, \tag{47}$$
where $\tau = t_c - t$, yielding model (28) in terms of $v$ as
$$\frac{\partial v}{\partial \tau} = 1 + \mathcal{M}[v], \tag{48}$$
where $\mathcal{M}[v]$ is a nonlinear operator on $v(x, \tau)$ that depends on the parameters $m$ and $n$.

We then seek a local solution to (48) in the form
$$v(x,\tau) = v_0(\tau) + \frac{X^2}{2}v_2(\tau) + \frac{X^4}{4!}v_4(\tau) + \cdots, \quad \text{as } X \to 0, \tag{49}$$
where $X = x - x_c$. This form assumes that the rupture profile is symmetric around the singularity location. Substitution of the expansion (49) into (48) and collecting coefficients in $X$ yield a coupled ODE system for $v_0$ and $v_2$. Note that the asymptotic similarity ansatz (10) implies that the local expansion around $x_c$ is
$$h(x,t) \sim \tau^\alpha H(X/\tau^\beta) = \tau^\alpha H(0) + \frac{X^2}{2}\tau^{\alpha-2\beta}H''(0) + \frac{X^4}{4!}\tau^{\alpha-4\beta}H^{(4)}(0) + O(X^6), \tag{50}$$
where $\alpha = 1/(m+5)$ from (42). This expansion should match with the local expansion of $h(x,t)$ for the transformed problem using (49),
$$h(x,t) = (\alpha^{-1}v(x,\tau))^\alpha = \alpha^{-\alpha}\left(v_0^\alpha + \frac{X^2}{2}\alpha v_2 v_0^{\alpha-1} + \frac{X^4}{4!}\alpha v_4 v_0^{\alpha-1} + O(X^6)\right). \tag{51}$$

Hence we have
$$H(0)\tau^\alpha = \alpha^{-\alpha}v_0^\alpha, \quad H''(0)\tau^{\alpha-2\beta} = \alpha^{1-\alpha}v_2 v_0^{\alpha-1}, \quad H^{(4)}(0)\tau^{\alpha-4\beta} = \alpha^{1-\alpha}v_4 v_0^{\alpha-1}. \tag{52}$$

Here we assume that $H(0), H''(0) = O(1)$ and $H^{(4)}(0) \ll 1$. Then (52) motivates the assumption (53) for $v_0$, $v_2$ and $v_4$:
$$v_0 = O(\tau), \quad v_2 = O(\tau^{1-2\beta}), \quad v_4 \ll v_0^{-1}v_2^2 \quad \text{for } \tau \to 0. \tag{53}$$
Under the conditions $n + m < 5$ and $m > -4$, we apply the scaling (53) as $\tau \to 0$ to reduce the coupled ODE system to
$$\frac{dv_0}{d\tau} = 1 + Ev_0^{2\beta-1}v_2, \quad \frac{dv_2}{d\tau} = Fv_0^{2\beta-2}v_2^2, \tag{54}$$
where
$$E = -4\alpha^{1-2\beta}, \quad F = -4\alpha^{1-2\beta}(2\alpha + 6\beta - 5).$$

In order to solve the nonlinear coupled ODEs (54) asymptotically as $\tau \to 0$, we consider three cases: $\beta < \frac{1}{2}$, $\beta = \frac{1}{2}$ and $\beta > \frac{1}{2}$. With $\beta$ defined by (42) and $m > -4$, these cases correspond to $n < 5$, $n = 5$ and $n > 5$, respectively. We will show that only the first case $n < 5$ serves as a valid restriction in addition to $0 < m + n < 5$ and $m > -4$ that allows a self-similar solution to (28).

To begin, for the case $\beta < \frac{1}{2}$ (that is, for $n < 5$), we seek $v_0$ and $v_2$ in forms of expansions in $\tau$ as $\tau \to 0$,

$$v_0 = A_0 \tau + A_2 \tau^{2-2\beta} + \cdots, \quad v_2 = B_0 \tau^{1-2\beta} + B_2 \tau^{2-4\beta} + \cdots, \tag{55}$$

where $A_0, A_2, B_0, B_2$ are constants to be determined. Substituting this assumption into (54), we can obtain the constants $A_0, B_0$ by solving the equations

$$-A_0 + 1 + E A_0^{2\beta-1} B_0 = 0, \quad B_0(2\beta - 1) + F A_0^{2\beta-2} B_0^2 = 0.$$

Taking the first two terms in $v_0(\tau)$ and $v_2(\tau)$ and using the scaling (42), we conclude from the expansion (51) that for parameters $(n, m)$ in the redefined region $(A)$, where

$$\text{Redefined region } (A) = \{(n, m) : 0 < n + m < 5, n < 5 \text{ and } m > -4\}, \tag{56}$$

the local form of $h(x, t)$ is given by

$$h(x,t) = \left(\frac{A_0}{\alpha}\right)^\alpha \tau^\alpha \left[\left(1 + \frac{\alpha B_0}{2A_0}\frac{X^2}{\tau^{2\beta}} + \cdots\right) + \frac{\alpha A_2}{A_0}\tau^{1-2\beta}\left(1 + \frac{B_2}{A_2}\frac{X^2}{\tau^{2\beta}} + \cdots\right) + \cdots\right]. \tag{57}$$

Accordingly, a revised bifurcation diagram on system parameters $(n, m)$ is provided in Figure 13.

In order to interpret the local structure (57), we consider the general change of variables

$$h(x,t) = \tau^\alpha \tilde{H}(\eta, s), \quad \tau = t_c - t, \quad \eta = X/\tau^\beta, \quad X = x - x_c, \quad s = -\ln \tau. \tag{58}$$

The change in temporal variable from $t$ to $s$ allows us to examine under what conditions the general solution $\tilde{H}(\eta, s)$ will approach a self-similar profile $H(\eta)$ as $s \to \infty$ (equivalent to $t \to t_c$ for finite-time rupture). With these similarity variables (58), the reduced PDE (28) transforms to the similarity equation

$$\frac{\partial \tilde{H}}{\partial s} - \alpha \tilde{H} + \beta \eta \frac{\partial \tilde{H}}{\partial \eta} = \mathcal{N}(\tilde{H}), \tag{59}$$

where the first three terms are obtained from the transformation of the time derivative, and $\mathcal{N}(\tilde{H})$ is a nonlinear operator describing the combination effects of evaporation, intermolecular forces and surface tension. The steady state of the similarity equation (59), $H(\eta)$, is the similarity profile defined in (10). To study the linear stability of the steady state $H(\eta)$ in the similarity variables, we consider solutions given by infinitesimal linear perturbations around these solutions,

$$\tilde{H}(\eta, s) = H(\eta) + \epsilon \sum_{k=1}^{\infty} \hat{H}_k(\eta) e^{\lambda_k s}. \tag{60}$$

Substitution of the form (60) into (58) with $s = -\ln \tau$ yields

$$h(x,t) = \tau^\alpha \left(H(\eta) + \epsilon \sum_{k=1}^{\infty} \hat{H}_k(\eta) \tau^{-\lambda_k}\right). \tag{61}$$

The definitions of similarity variables in (10) rely on knowledge of exact rupture time $t_c$ and singularity position $x_c$, which are continuously dependent on initial conditions. As described in [53], shifts in time around $t_c$ and in space around $x_c$ can lead to positive symmetry eigenmodes in linear stability analysis of the self-similar solutions. Specifically, when errors in the estimate of $t_c$ and $x_c$ are included, the temporal and spatial translation variables take the forms

$$\bar{\tau} = t_c - t - C_T \delta, \quad \bar{X} = x - x_c - C_X \delta, \quad \bar{\eta} = \bar{X}/\bar{\tau}^\beta. \tag{62}$$

Substituting (62) into (10) we obtain as $\epsilon \to 0$,

$$h(x,t) = \bar{\tau}^\alpha \left(H(\bar{\eta}) + \delta \bar{\tau}^{-1} C_T \left(\alpha H(\bar{\eta}) - \beta \bar{\eta} H'(\bar{\eta})\right) + \delta \bar{\tau}^{-\beta} C_X H'(\bar{\eta}) + O(\delta^2)\right). \tag{63}$$

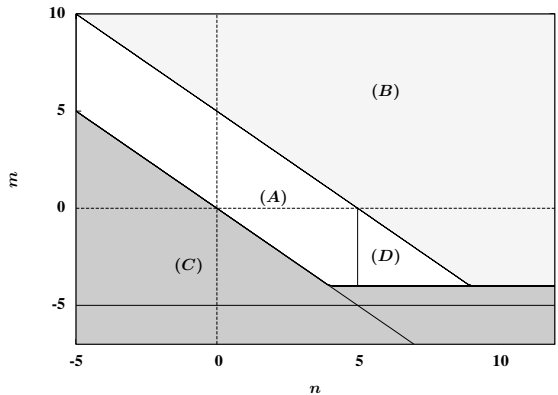

Figure 13: Bifurcation diagram on system parameters $(n, m)$: $(A)$ is for second-order rupture, $(B)$ for fourth-order rupture, $(D)$ for non-self-similar second-order rupture, $(C)$ for uniform thinning regime. In particular, the horizontal line $m = -5$ separates the finite-time and infinite uniform thinning region, $(C_1)$ and $(C_2)$, while the extended line $m + n = 0$ in $(C)$ is a threshold that differentiates spatial variation behaviors described by (39).

The $O(\delta)$ contribution in (63) corresponds to the two symmetry modes $\lambda_T = 1$ and $\lambda_X = \beta$ due to the invariance of temporal and spatial translations of the rupture singularity, and can be derived from the local expansion analysis by allowing singular solutions in (55). In the above local expansion procedure, we implicitly assume in (55) that correct values of $t_c$ and $x_c$ are selected, which gives us $C_T = C_X = 0$. Consequently, the positive symmetry modes $\lambda_T$ and $\lambda_X$ are not involved in the expansion (57).

Since the local structure (57) is consistent with the expansion (61), we conclude that the asymptotic similarity assumption (10) is valid for $\beta < \frac{1}{2}$ (the $n < 5$ case). Also the order of the correction term in (57) suggests that all the eigenvalues $\lambda_k$ other than the symmetry modes $\lambda_T$ and $\lambda_X$ in (60) are negative, which implies that the profile $H(\eta)$ is a stable steady state of the time-dependent similarity equation (59).

3.2.2. Region $(D)$: non-self-similar rupture

Next we asymptotically solve the coupled ODEs (54) for the case $\beta \geq \frac{1}{2}$, that is, $n \geq 5$. We will show that this corresponds to a non-self-similar rupture regime where $m$ and $n$ are in region $(D)$ in Figure 13 defined by

$$(D) = \{(n, m) : n + m < 5, n \geq 5, \text{ and } m > -4\}. \tag{64}$$

Assume that $v_0$ and $v_2$ satisfy

$$E v_0^{2\beta-1} v_2 \ll 1 \quad \text{for } \tau \to 0. \tag{65}$$

As we are interested in the form of $v_0(\tau)$ with $v_0(0) = 0$, after reducing the first equation in (54) to $v_0' - 1 = 0$, we have $v_0 \sim \tau$. Then the differential equation for $v_2$ becomes

$$\frac{dv_2}{d\tau} = F \tau^{2\beta-2} v_2^2.$$

Integration of this equation leads to the following two cases asymptotically as $\tau \to 0$:

$$v_2(\tau) = c_1 + \frac{c_1^2 F}{2\beta - 1} \tau^{2\beta-1} + O(\tau^{4\beta-2}), \qquad \beta > \frac{1}{2}, \tag{66a}$$

$$v_2(\tau) = \frac{1}{F|\ln \tau|} + \frac{c_2}{F|\ln \tau|^2} + O\left(\frac{1}{|\ln \tau|^3}\right), \qquad \beta = \frac{1}{2}, \tag{66b}$$

for constants $c_1, c_2$ that depend on initial conditions. From (66) we observe that for $\beta \geq \frac{1}{2}$ the assumptions (65) and (53) are satisfied. Substitution of the expansion (66) into the ODE for $v_0$ in (54) yields

$$v_0(\tau) = \tau + \frac{c_1 E}{2\beta} \tau^{2\beta} + O(\tau^{4\beta-1}), \qquad \beta > \frac{1}{2}, \tag{67a}$$

$$v_0(\tau) = \tau + \frac{E\tau}{F|\ln \tau|} + O\left(\frac{\tau}{|\ln \tau|^2}\right), \qquad \beta = \frac{1}{2}. \tag{67b}$$

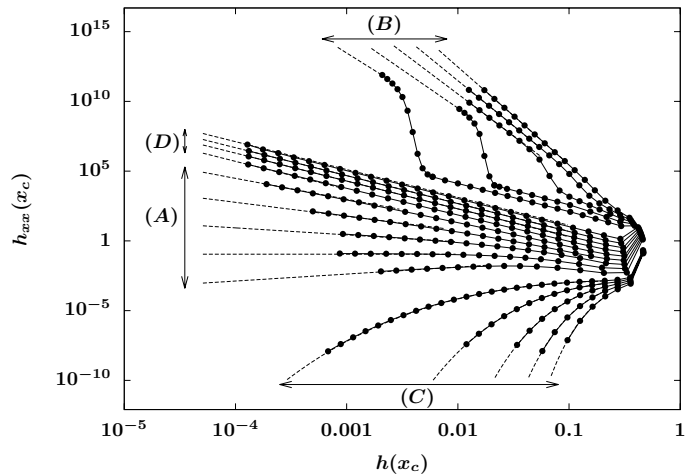

Figure 14: Numerical verification of solution behaviors starting from $h_0(x) = 0.5 - 0.1\exp(-100(x - L/2)^2)$ for $0 \leq x \leq 10$ with fixed $m = -2$ and over a range of $n$ in different parameter regions. Region $(A)$: $n = 2.5, 3, \cdots, 4.5$; Region $(B)$: $n = 8, 8.5, \cdots, 10$; Region $(C)$: $n = 1, 1.2, \cdots, 1.8$; Region $(D)$: $n = 5, 5.5, 6, 6.5$. Numerical results are represented by dots, and analytic predictions (31) are plotted in dashed lines with the scaling parameter $\nu$ given by (43) for region $(A)$, (73) for region $(B)$, (39c) for region $(C)$ and (69) for region $(D)$.

With the expansion (51) and the leading order terms in (67) and (66) the corresponding local expansions for $h(x, t)$ can be written as

$$h(x,t) = \alpha^{-\alpha}(t_c - t)^\alpha \left(1 + D_2 \frac{(x-x_c)^2}{(t_c-t)} + D_0(t_c-t)^{2\beta-1} + \cdots\right), \qquad n > 5, \qquad (68a)$$

$$h(x,t) = \alpha^{-\alpha}(t_c - t)^\alpha \left(1 + \frac{\alpha E}{F|\ln(t_c-t)|} + \frac{\alpha(x-x_c)^2}{2F(t_c-t)|\ln(t_c-t)|} + \cdots\right), \quad n = 5, \quad (68b)$$

where the coefficients $D_0$ and $D_2$ depend on the initial condition $h_0(x)$. The inconsistency between (68) and the self-similar expansion (50) implies that for $n \geq 5$, PDE (28) does not have an asymptotic self-similar solution that strictly follows (10). In particular, recalling that $\alpha = 1/(m+5)$, for $n > 5$ the asymptotic form (68a) suggests that the solution to model (28) asymptotically satisfies (31) and (32) with

$$\nu = -m - 4, \quad \mu = -m - 4, \tag{69}$$

which is different from the relationship (43) derived for the self-similar case. For $n = 5$ and $m = -2$, the asymptotic form in (68b) is consistent with the ansatz derived in [30] for equation (21) with $\sigma = 1, \rho = -2$, which is identical to the leading order PDE (46) from dimensional analysis up to a constant multiple.

In order to verify the analytical results above, we implemented a sequence of numerical PDE simulations with fixed $m = -2$ and varying $n$ in regions $(A)$, $(B)$, $(C)$ and $(D)$. A comparison of the relationships between $h_{xx}(x_c, t)$ and $h(x_c, t)$ is plotted in Figure 14. As $h(x_c)$ approaches zero, simulations show that $h_{xx}(x_c, t)$ and $h(x_c, t)$ satisfy (43) for $0 < m + n < 5$ and $n < 5$ in region $(A)$. While finite-time rupture phenomenon usually happens with $h_{xx}(x_c, t)$ approaching infinity, the positive slope of the bottom line in region $(A)$ in Figure 14 indicates that $h_{xx}(x_c, t) \to 0$ as $t \to t_c$ can happen. Specifically, for $0 < m+n < 1$, we have $\alpha, \beta > 0$ and $\alpha - 2\beta > 0$. Therefore using (30) we see that the self-similar solution is focusing around the critical location $x_c$ while $h_{xx}(x_c, t)$ approaches zero as the rupture develops. The rupture solutions in this case are similar to the one depicted in Figure 12, but are flatter locally around $x_c$.

For $0 < m+n < 5$ and $n > 5$ in case $(D)$ the slope of parallel lines agrees with (69) and the intercepts of the lines depend on initial data, which confirms that the rupture solutions in this regime does not approach a self-similar profile. For $m+n < 0$ in case $(C)$, uniform thinning phenomenon occurs following (39c) with initial spatial perturbations damping out very quickly. With $m + n > 5$ in region $(B)$, we observe that in the early stage, the solution evolves following the linear stability result (39a). For $h(x_c)$ sufficiently small, the dynamics is in the localized rupture stage, and the relationship between $h_{xx}(x_c)$ and $h(x_c)$ satisfies (31) with the scaling parameter given by (73). Spatial oscillations are observed between the two stages, and correspond to the transition from a flatter slope to a steeper one in Figure 14. This is one of the interesting features of fourth order rupture solutions and we will present our results for this regime in the next section.

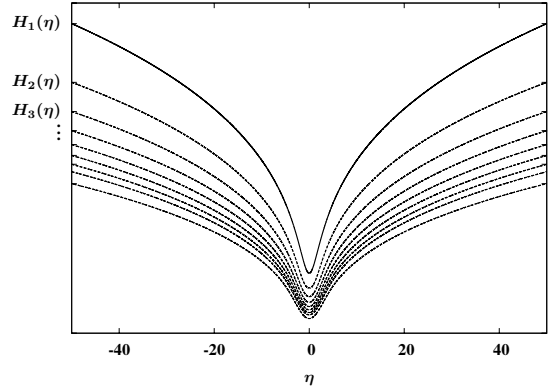

Figure 15: Fourth-order similarity solutions $H_k(\eta)$, $k = 1, 2, 3, \cdots$, of (71) for $n = 6, m = 1$.

To summarize, for $(n, m)$ in the redefined region $(A)$, self-similar rupture solutions to model (28) are governed by the second-order PDE (46). For $(n, m)$ in region $(D)$ we discovered a family of non-self-similar rupture solutions. In the next section we focus on asymptotically self-similar solutions to model (28) that are governed by a fourth-order equation.

3.3. Region $(B)$: Fourth order rupture

The final region in the parameter plane for (28) is region $(B)$:

$$(B) = \{(m, n) : m + n > 5 \text{ and } m > -4\}. \tag{70}$$

Like the case for Region $(A)$, for $m + n > 5$ there are also two possible dominant balances in (29). One of them is

$$-\alpha H + \beta \eta H_\eta + \frac{1}{H^{4+m}} + \frac{H_{\eta\eta}}{H^m} = 0,$$

obtained with the scaling exponents, $\alpha = 1/(m+5)$ and $\beta = 5/[2(m+5)]$. However, this balance will not be applicable because it corresponds to an illposed PDE,

$$h_t = -h^{-m} h_{xx} - h^{-4-m}.$$

Consequently, the dominant balance for rupture solutions in region $(B)$ as $\tau \to 0$ is

$$O\left(\tau^{\frac{1-2m-n}{2m+n}}\right): \quad -\alpha H + \beta \eta H_\eta + \frac{H_{\eta\eta}}{H^m} + (H^n H_{\eta\eta\eta})_\eta = 0, \tag{71}$$

determined by the scaling exponents,

$$\alpha = \frac{1}{n + 2m}, \quad \beta = \frac{n + m}{2(n + 2m)}. \tag{72}$$

Recalling (31, 32), these $(\alpha, \beta)$ yield

$$\nu = 1 - n - m, \quad \mu = 1 - n - 2m, \tag{73}$$

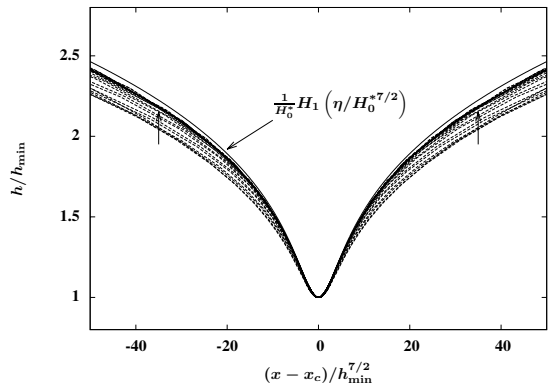

Figure 16: Rescaled solution profiles from a numerical simulation of (28), with $m=1, n=6$ and initial condition $h_0(x) = 0.5 + 0.1\cos(4\pi x)$ on $0 \leq x \leq 1/2$, showing convergence to the similarity solution (solid line) scaled as $H_1\left(\eta/(H_0^*)^{7/2}\right)/H_0^*$ of the leading order ODE (71), where $H_0^* = H_1(0) = 0.7997$.

and going back to Figure 8, at $m=1$, these indeed give $\nu = -n$ and $\mu = -n-1$ as observed in the PDE simulations, as well in other previous figures (Figs. 9(left) and 14). Similarity equation (71) corresponds to the fourth-order self-invariant leading order PDE,

$$\frac{\partial h}{\partial t} = -\frac{\partial}{\partial x}\left(h^n \frac{\partial^3 h}{\partial x^3}\right) - \frac{1}{h^m}\frac{\partial^2 h}{\partial x^2}, \tag{74}$$

which clarifies the nature of the rupture singularity for (28) in region $(B)$ as stemming from a balance between the regularizing surface tension term and the destabilizing second order operator.

Solving the fourth-order nonlinear ODE (71) subject to the far field Robin boundary condition (44) numerically, we obtain a discrete family of similarity solutions, $H_k(\eta)$ for $k = 1, 2, 3, \cdots$, see Figure 15. Similar behavior for similarity solutions of other fourth-order rupture problems was observed in [8, 53]. As shown in Figure 16, numerical simulations of PDE (28) with $n = 6, m = 1$ show scaled profiles converging to the primary self-similar solution $H_1(\eta)$ as $h_{\min} \to 0$. Indeed applying linear stability analysis, as described in [52, 53], to the similarity solutions $H_k(\eta)$ as steady states of (59) shows that only the $k = 1$ rupture solution is stable in region $(B)$.

It is notable that although equation (74) captures the asymptotic features of rupture in model (28), the transient analysis of disturbances to flat films differs somewhat from (35b). Dropping the evaporation term $1/h^{4+m}$ from the full model (28) to get the reduced PDE (74) results in flat films, $h = \bar{h}$, being steady states rather than having evolving thicknesses, $h = \bar{h}(t)$ as in (36); this is observed in the far-field film heights (away from the rupture at $x_c = 2.5$) in the two simulations shown in Figure 17. Approximating the solution as $h(x,t) = \bar{h} + \delta e^{ikx} e^{\lambda t} + O(\delta^2)$, and applying linear stability analysis, we obtain the dispersion relation for (74) at $O(\delta)$ as

$$\lambda = \frac{k^2}{\bar{h}^m} - k^4 \bar{h}^n, \tag{75}$$

showing the model to be longwave-unstable for wavenumbers $0 < k < \bar{h}^{-(n+m)/2}$. This is comparable to equation (35b) for (28). Indeed Figure 17 shows that starting from the same initial Gaussian perturbation, both equations exhibit equivalent spatial oscillations until nonlinear effects lead to rupture at a local minimum. Similar dynamics involving a transition from linearized behavior to fourth-order rupture were also shown in Figure 9, though there spatial oscillations were less evident due to the domain size and other parameter choices. A more detailed study of the variety of dynamics stemming from the transient behaviors will be given in [32].

The influence of surface tension, as represented in the underlying thin film equation,

$$\frac{\partial h}{\partial t} = -\frac{\partial}{\partial x}\left(h^n \frac{\partial^3 h}{\partial x^3}\right), \tag{76}$$

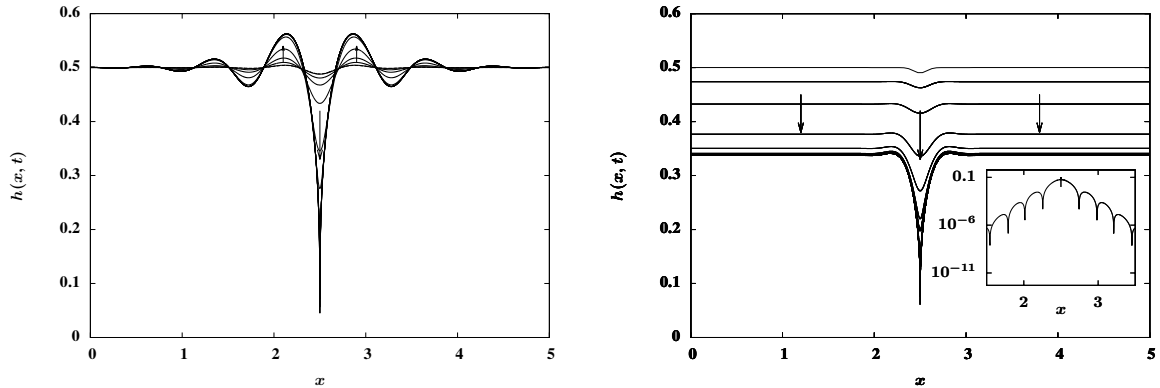

Figure 17: (Left) Dynamics leading to rupture in the leading order equation (74) with $n = 6, m = 1$ starting from the initial condition $h_0(x) = 0.5 - 0.01 \exp(-100(x - 5/2)^2)$. (Right) Corresponding dynamics in the full model (28). The inset shows a plot of $|h(x,t) - \bar{h}(t)|$ at a time $t = 0.0021392$ (near rupture) to more clearly display the presence of spatial oscillations in the film height.

that is present in (74) has been the focus of extensive PDE analysis over recent decades. The seminal results of Bernis and Friedman [6] established the preservation of nonnegativity for solutions of (76) for $n \geq 4$ on bounded domains with no-flux boundary conditions. Most of our focus in this paper has been in examining how non-conservative (evaporative) effects can produce rupture to overcome the intermolecular forces that would otherwise preserve positivity. However, as shown in Figures 17 for $n > 4$ in (74), non-conservative effects can also overcome surface tension to yield finite-time singularities.

## 4. Conclusions

Finite-time rupture solutions of a fourth-order parabolic nonlinear thin film equation with a non-conservative term (3) have been analyzed. Unlike the model without evaporation effects ($\gamma = 0$) where the positivity of solutions is guaranteed, evaporative fluxes can overcome intermolecular forces and lead to self-similar singularities. In contrast to rupture solutions of many other thin film equations, the evaporation term in our model controls the similarity scaling exponents. Since the surface tension is negligible, self-similar rupture solutions are asymptotically governed by a second-order fast diffusion type equation with a singular absorption term.

We have also explored distinctive evaporation-driven dynamic behaviors of a general model (28) with different mobility coefficients. More precisely, in addition to the second-order self-similar rupture observed in model (3), uniform thinning with damping spatial variation and another two types of finite-time singularities can occur with system parameters in different ranges (see Figure 13). Besides the separation between second-order and fourth-order self-similar rupture solutions in the bifurcation diagram that has been derived via formal dimensional analysis, we have also identified a family of non-self-similar rupture through a local expansion method.

A number of interesting questions regarding the models (3) and (28) remain to be solved. First, we have simplified the analysis of (3) for the finite-time rupture behavior by selecting $P_0$ such that $\Pi(h) < 0$, which guarantees the monotonic decrease of the mean height in time. For $P_0$ in other scenarios, as is discussed in [32], the dynamic evolution becomes more complicated, since more transient stages between various non-trivial steady states to the model will be involved.

Inspired by the early and later stage scaling transitions (Figure 6) in the evaporation model (12), we are also interested in exploring other possible transitions. For example, if the mobility function in (12) is changed from $h^3$ to $h^{4.5}$, then similar analysis as in section 2.1 and 2.2 will lead to two approximations for the evaporative term in the early and later stage of the rupture evolution, which correspond to a transition from $(n, m) = (4.5, 1)$ to $(n, m) = (4.5, 0)$ in the generalized model (28). Unlike the transition in Figure 6 where both stages belong to the second-order rupture regime, the dynamic evolution of this equation is expected to involve a transition from the fourth-order rupture, the case $(B)$, to the second-order rupture, the case $(A)$. Further investigation is needed to explore the influence on scaling transitions of more general forms of the evaporation term, for instance $J = -\gamma p(h)/(h + K_0)^m$.

For the general model (28), our investigation of the finite-time singularities has been based on asymptotic analysis accompanied by adaptive PDE simulations with high resolution. Therefore many of the statements about the dynamics made here can be taken as conjectures on this PDE that are open problems motivating further work in rigorous analysis. For instance, as numerical evidence suggests that finite-time rupture occurs for general initial conditions with system parameters $(n, m)$ outside of region $(C)$, we may ask: does any solution with nontrivial initial conditions have global existence? Moreover, with $(n, m)$ in region $(B)$, it is observed numerically that the rescaled PDE solution asymptotically approaches to a fourth-order similarity ODE solution. However, rigorous theory is still needed to show the existence of these ODE solutions [25, 26], and whether the stable similarity solution attracts all nontrivial initial conditions.

The dynamics of the model in critical cases with $(n, m)$ on the thresholds $m + n = 5$, $m + n = 0$, $n = 5$ and $m = -4$ would be other interesting and challenging problems. For $n = 5$ at the boundary of region $(D)$, we have derived a local expansion formula for rupture solutions around the singularity location in section 3.2.2 using the method from [30]. Refined analysis will be needed for the other thresholds. In addition, analytical descriptions beyond the linear stability analysis of transient behaviors before the final stage rupture also remain to be addressed. For example, as $(n, m)$ approaches the threshold $m = -4$ with $n > 9$ between region $(B)$ and $(C)$ from the fourth-order rupture region, Figure 9 suggests that the length of the transient behaviors prior to the self-similar fourth-order rupture approaches infinity and further analysis of this limit is needed.